# Using Multi-Threshold Threshold Gates in RTD-based Logic Design. A Case Study

Héctor Pettenghi, María J. Avedillo, and José M. Quintana

Instituto de Microelectrónica de Sevilla, CNM, Sevilla, SPAIN
E-mail: {hector, avedillo, josem}@imse.cnm.es

*Abstract* - **The basic building blocks for Resonant Tunnelling Diode (RTD) logic circuits are Threshold Gates (TGs) instead of the conventional Boolean gates (AND, OR, NAND, NOR) due to the fact that, when designing with RTDs, threshold gates can be implemented as efficiently as conventional ones, but realize more complex functions. Recently, RTD structures implementing Multi-Threshold Threshold Gates (MTTGs) have been proposed which further increase the functionality of the original TGs while maintaining their operating principle and allowing also the implementation of nanopipelining at the gate level. This paper describes the design of *n*-bit adders using these MTTGs. A comparison with a design based on TGs is carried out showing advantages in terms of latency, device counts and power consumption.**

## I. INTRODUCTION & BACKGROUND

Resonant tunnelling diodes (RTDs) are very fast non linear circuit elements which exhibit a negative differential resistance (NDR) region in their current-voltage characteristics (see Figure 1a) which can be exploited to significantly increase the functionality implemented by a single gate in comparison to conventional technologies, thus reducing circuit complexity [1]. Circuit applications of RTDs are mainly based on the MOnostable-BIstable Logic Element (MOBILE) [2], [3]. The MOBILE, shown in Fig. 1a, is a rising edge triggered current controlled gate which consists of two RTDs connected in series and driven by a switching bias voltage $V_{bias}$. When $V_{bias}$ is low both RTDs are in the on-state (or low resistance state) and the circuit is monostable. Increasing $V_{bias}$ to an appropriate maximum value ensures that only the device with the lowest peak current switches (quenches) from the on-state to the off-state (the high resistance state). Output is high if the driver RTD is the one which switches and it is low if it is the load. Peak currents are proportional to RTDs areas. Assuming equal current densities, the smallest RTD is the one which switches. Logic functionality is achieved by embedding an input stage which modifies the peak current of one of the RTDs. An input stage com-posed of the series connection of an RTD and a FET transistor has been proposed [4]. This is the idea of the inverter MOBILE shown in Fig. 1b. Only when a logic one is applied to the gate of a transistor its associated RTD contributes to the $NDR_2$ current. RTDs areas are selected such that the relation of the peak currents of $NDR_1$ and $NDR_0$ depends on whether the external input signal $V_{in}$ is "1" or "0". Transistor are sized such that RTDs are the current limiting devices in the series connection of the RTD and the transistor. For $V_{bias}$ high the output node maintains its value even if the input changes. That is, this circuit structure is self-latching allowing to implement pipeline at the gate level without any area overhead associated to the addition of the latches which allows very high throughtoutput.

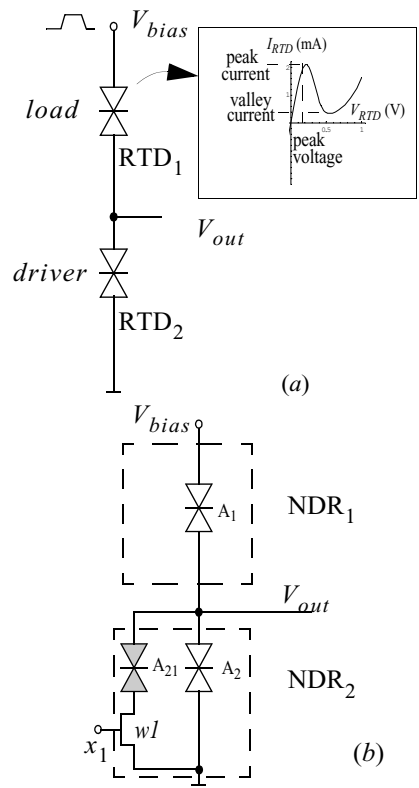

Figure 1. MOBILE circuits, a) basic MOBILE, b) MOBILE inverter.

1. This effort was partially supported by the Spanish Government under project TEC2004-02948/MIC.

Threshold logic [5] is a computational model widely used in the design of RTD based circuits [4]. That is, the basic building blocks for RTD logic circuits are Threshold Gates (TGs) instead of the conventional Boolean gates (AND, OR, NAND, NOR). This is due to the fact that when designing with RTDs, threshold gates can be implemented as efficiently, in terms of performance and complexity, as conventional Boolean gates but realize more complex functions.

A TG or linear separable function is defined as a logic gate with $n$ binary input variables, $x_i, (i = 1, ..., n)$, one binary output $y$, and for which there is a set of $(n+1)$ real numbers: threshold $T$ and weights $w_1, w_2, ..., w_n$, such that its input-output relationship is defined as $y = 1$ iff $\sum_{i=1}^{n} w_i x_i \geq T$, and $y = 0$ otherwise. Sum and product are the conventional, rather than the logical, operations. The set of weights and threshold can be denoted in a more compact vector notation way by $[w_1, w_2, ..., w_n; T]$.

TGs are efficiently realized by resorting to the operation principle of the clocked series connection of a pair of RTDs (MOBILE). The functionality of the MOBILE inverter can be extended to the TG as depicted in Figure 2 showing the circuit structure for a generic threshold gate of two inputs $[w_k, w_l; T]$, where $w_k$ represents a positive weight and $w_l$ represent a negative one [4]. The areas $A_1$ and $A_2$ are determinated by the threshold to be implemented, and the areas $A_u$ and $B$ are selected according to the technology and to the required design trade-offs involving speed, power and robustness against parameter variations. Threshold gates with a larger number of inputs can be easily realized extending this topology.

More recently, RTD structures implementing Multi-Threshold Threshold Gates (MTTGs) [6] have been proposed which further increase the functionality of the original TGs while maintaining their MOBILE operating principle and associated advantages [7], [8], [9], [10].

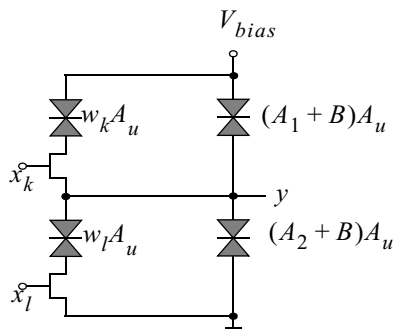

Figure 2.- Circuit structure for $[w_k, w_l; T]$.

MTTGs are a generalization of the conventional TGs. Formally, A $k$-threshold MTTG is a logic element with $n$ binary input variables, $x_i, (i = 1, ..., n)$, one binary output $y$, and for which there is a set of $(n+k)$ real numbers: thresholds $T_i, (i = 1, ..., k)$, and weights $w_1, w_2, ..., w_n$, such that its input-output relation is defined as: $y = 1$ iff 

$$T_{2j-1} \leq \sum_{i=1}^{n} w_i x_i < T_{2j}, \text{ with } T_{j+1} > T_j, (j = 1, ..., k/2);$$

output $y$ is equal to zero otherwise. As in the TGs, the set of weights and thresholds can be denoted in a more compact vector notation way by $[w_1, w_2, ..., w_n; T_1, ..., T_k]$.

There are a number of theoretical properties for MTTGs which suggest they can be extremely powerful in digital design [6]. In particular, any $n$-input switching function can be realized by a single MTTG with at most $(2^n -1)$ thresholds simply assigning $w_i = 2^i - 1$. Moreover, usually exponential weights and exponential number of thresholds are not required. Thus, from a logic point of view, the MTTG is a powerful concept.

This paper describes the design of $n$-bit adders using RTD-based MTTGs. The rest of the paper is organized as follows. Section II describes an adder implemented by means of TGs and reported by Pacha et al [4]. Section III describes the proposed adder using MTTGs and the realization of tthese more powerful gates on the basis of the concept of the MOBILE. Section IV compares three-bit adders designed with each of the architectures.

## II. ADDERS BASED ON TGS

Figure 3a shows the logic diagram of a nanopipelined carry propagation $n$-bit adder based on TGs reported in [4]. It is comprised of the chain of full adders (FAs) and MOBILE followers (latches) to support pipeline. The first FA has been simplified. Since the carry output of the FA is a threshold function, it is implemented by a single threshold gate. The sum output of a FAs is a three-input EXOR function, which is not linear separable, and thus require a network of TGs to be realized. It is implemented using a two stage network with four gates as depicted in Figure 3a. Figure 3b-e depict the TG based realizations for the two-input EXOR (Figure 3b), the FA (Figure 3c), the two-input AND (Figure 3d) and latches (Figure 3e). Bias signals to operate cascaded MOBILE-type circuits are also shown (Figure 3f). A four phase (evaluation, hold, reset and wait) overlapping clocking scheme is used. Note that second stage ($V_{bias2}$) evaluates while the first stage ($V_{bias1}$) is in the hold phase. For a number of logic levels greater than three, four bias signals are required. In one clock period all the gates are activated. Data can be processed at a frequency determined by the operation speed of four chained MOBILE gates.

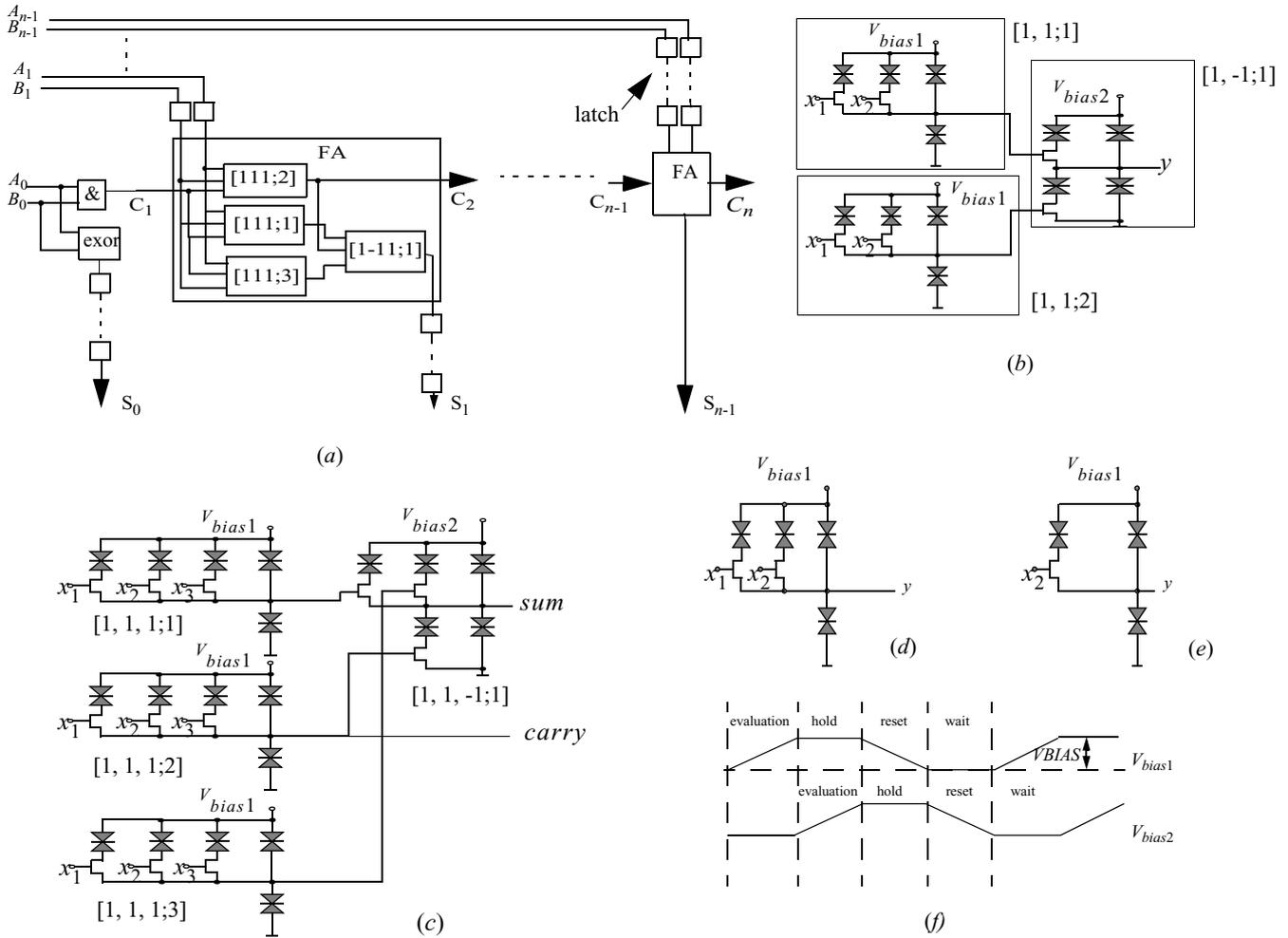

Figure 3. *n*-bit adder based on TGs. (a) logic diagram. (b) RTD-based realization of two-input EXOR. (c) RTD-based realization of FA. (d) RTD-based realization of two-input AND. (e) Latch. (f) Bias scheme for MOBILE circuits.

### III. ADDERS BASED ON MTTGS

Figure 4a depicts the circuit topology we have proposed for a generic MTTG with positive weights defined by $[w_1, w_2; T_1, T_2]$, i.e., two inputs and two thresholds [8]. The specific function implemented by such a circuit depends on the areas of the RTDs. The concept of RTD-based MOBILE is extended to a circuit consisting of three or more RTDs in series. The switching sequence in series-connected RTDs begins also with the RTD with the smallest peak current [11]. If this peak current can be controlled by external inputs, this sequence can be varied and different functions can be obtained at the output nodes.

Figure 4b depicts a more efficient alternative realization of two input two threshold MTTGs [10]. Circuit in Figure 4a operates on the basis of current comparison: the NDR with the lowest peak current switches off. Inputs modulate the peak currents of both the bottom ($NDR_2$) and the middle ($NDR_1$) NDRs, although by a different amount since the areas of the RTDs of their associated input stages are different. Thus, concerning $NDR_1$ and $NDR_2$, and assuming, without loss of generality, $A_{11} > A_{21}$ ($A_{12} > A_{22}$), the same result of the current comparison could be obtained if only $NDR_1$ is modulated with input stages with area $A_{11} - A_{21}$ ($A_{12} - A_{22}$) for input variable $x_1$ ($x_2$). That is what has been done in Figure 4b. However, area relationships of $NDR_1$ and $NDR_2$ with $NDR_0$ must be kept for circuit in Figure 4b to implement the same functionality that circuit in Figure 4a. To achieve this, input stages of area equal to $A_{21}$ ($A_{22}$) for $x_1$ ($x_2$) common to both have been added to the proposed circuit. Note that the area of some RTDs is reduced with respect to the solution in Figure 4a, thus allowing to reduce the width of their associated transistors.

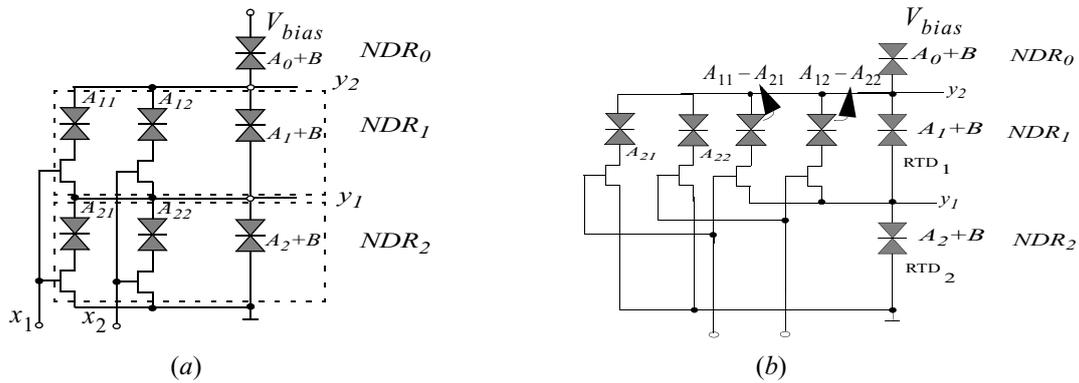

Figure 4.- Circuit structures for generic two-input two-threshold MTTG $[w_1, w_2; T_1, T_2]$ with positive weights.

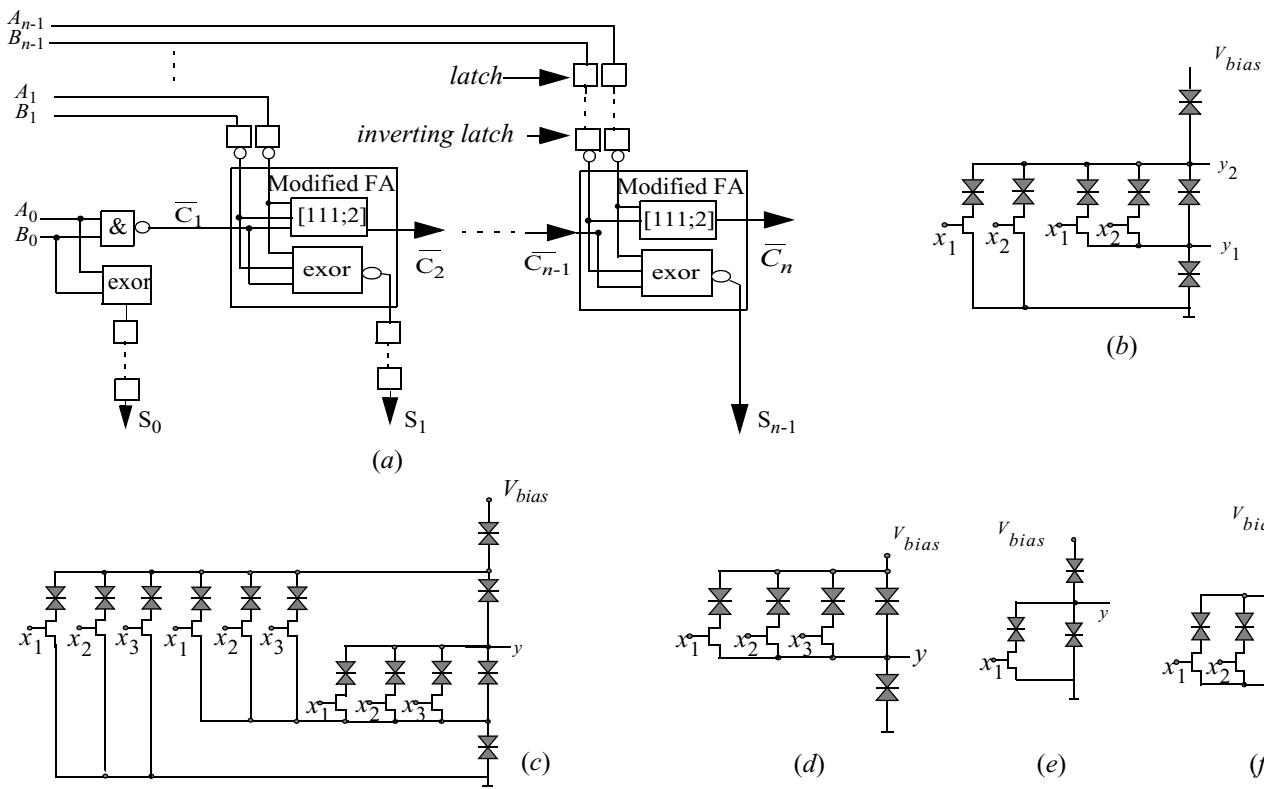

Figure 5. $n$-bit adder based on MTTGs. (a) logic diagram. (b) RTD realization of two-input EXOR. (c) RTD realization of three-input EXNOR. (d) RTD realization of Majority [111;2]. (e) inverting latch. (f) RTD realization of two-input NAND.

Figure 5a depicts the logic diagram of the proposed nanopipelining carry-propagation $n$-bit adder. Note that it uses modified full adders in which complemented inputs are processed. Each of them comprises only two logic gates (only one logic level): the three-input mayority and the three-input EXNOR. The first one is a TG ([1, 1, 1; 2]) and the second one is an MTTG which can be implemented by a single RTD gate extending the topology above described to four RTDs in series. The FA modification has been done to avoid using a three input EXOR which can not been implemented with the MTTG topology described. Additional latches (implemented as MOBILE buffers) and inverting latches (implemented as MOBILE inverters) are also required to support pipeline. Due to the fact that the proposed adder needs complemented inputs in the modified FAs, it is required to use an inverter in the last stage of the input latches.

Note that he circuit showed in Figure 3a requires a two-level network for the full adder while the architecture in Figure 5a implements the modified full adder with a single level of gates. This means the latency of the proposed adder is reduced.

IV. COMPARISON

Three-bit adders have been designed to operate up to 750MHz using each of the architectures in a non commercial but university InP technology in which RTD and transistors can be co-integrated [12]. Minimization of power has been the design target. Simulation results at the operating frequency of 750MHz are depicted in Figure 6. Figure 6a shows the results for the previously reported adder, Figure 6b shows the results for the proposed gate. Note that in this figure the carry is complemented although it can be obtained without complementation changing the majority gate [1, 1, 1; 2] of the last stage to [-1, -1, -1; -1] without increasing of the latency.

For three bit adders operating at 750 MHz the design based in TGs has 15% more devices, its latency is 33% larger and consumes 20% more power than the proposed in this work based on MTTGs.

V. CONCLUSIONS

In this paper a novel nanopipelined $n$-bit adder based on RTDs and HFET devices has been described. A comparative analysis has been carried between this proposed design and an adder previously reported. The proposed design has several advantages: lower number of transistors, smaller total area of RTDs, lower latency and lower power consumption. Nanopipelined architectures for multipliers and divisors recently reported [13] can take advantage of these advanced proposed adders.

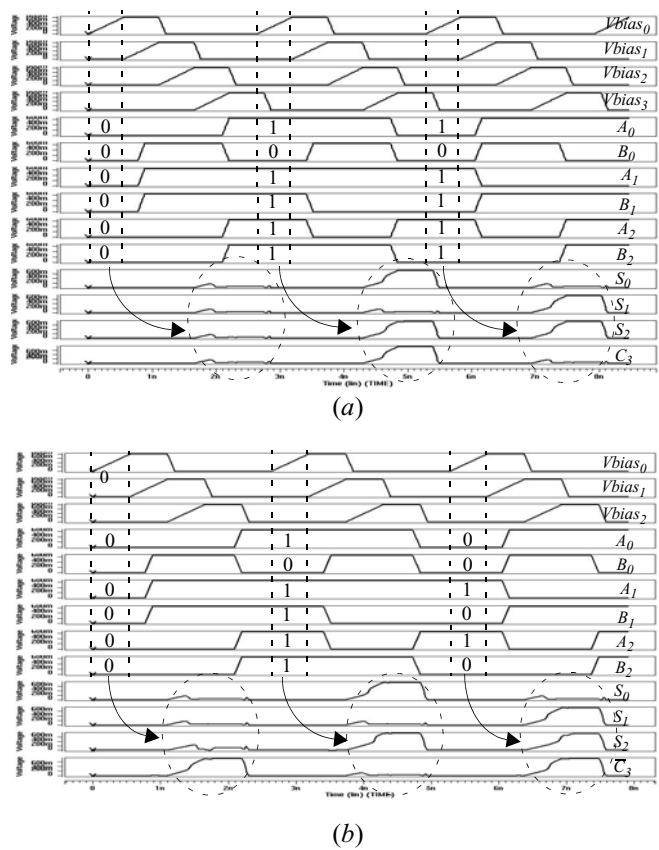

Figure 6.- Simulations results for three-bit adders. (a) TG based. (b) MTTG based.